\documentclass[conference,10pt]{IEEEtran}
\ifCLASSINFOpdf
\else
\fi
\ifCLASSINFOpdf
\else
\fi
\usepackage{latexsym}
\usepackage{amsmath}    
\usepackage{verbatim}   
\usepackage{color}      
\usepackage{setspace}
\usepackage{rotating}

\usepackage{epsfig}
\usepackage{graphicx}
\usepackage{amssymb}
\usepackage{amsthm}
\usepackage[acronym,toc,shortcuts]{glossaries}
\usepackage{multirow}
\usepackage{caption}
\usepackage{subcaption}
\usepackage{algpseudocode}
\usepackage{color}
\usepackage[table,xcdraw]{xcolor}
\usepackage{soul}
\usepackage{url}
\usepackage{balance}
\usepackage{cite}
\usepackage{enumitem}

\usepackage{algorithm}
\usepackage{algorithmicx}
\usepackage{longtable}
\usepackage{float}
\usepackage{array}
\usepackage{supertabular}
\usepackage[bookmarks=false]{hyperref} 
\usepackage{csquotes}
\usepackage{amsmath}

\usepackage{makecell}

\makeglossaries
\newacronym{2D}{2D}{two dimensional}
\newacronym{3D}{3D}{three dimensional}
\newacronym{3GPP}{3GPP}{The 3rd Generation Partnership Project }
\newacronym{5G-PPP}{5G-PPP}{5G Infrastructure Public Private Partnership}
\newacronym{4D}{4D}{four dimensional}
\newacronym{AAA}{AAA}{Authentication, Authorization and Accounting}
\newacronym{ABSF}{ABSF}{Almost-Blank Subframe}
\newacronym{AHP}{AHP}{Analytical Hierarchical Process}
\newacronym{AP}{AP}{Access Point}
\newacronym{API}{API}{Application Programming Interface}
\newacronym{APN}{APN}{Access Point Name}
\newacronym{AR}{AR}{Augmented Reality}
\newacronym{AWGN}{AWGN}{additive white Gaussian noise}
\newacronym{BBU}{BBU}{baseband unit}
\newacronym{BE}{BE}{best-effort}
\newacronym{BET}{BET}{Blind Equal Throughput}
\newacronym{BLAST}{BLAST}{Bell Laboratories Layered Space-Time}
\newacronym{BS}{BS}{Base Station}
\newacronym{BSS}{BSS}{Business Support System}
\newacronym{BTP}{BTP}{Backhaul Transport Provider}
\newacronym{CA}{CA}{carrier aggregation}
\newacronym{CCAS}{CCAS}{cooperative collision avoidance system}
\newacronym{CELL-ID}{CELL-ID}{cell identification ID}
\newacronym{CoV}{CoV}{Coefficient of Variation}
\newacronym{CP}{CP}{cyclic prefix}
\newacronym{CPU}{CPU}{central processing unit}
\newacronym{CoMP}{CoMP}{Coordinated Multipoint}
\newacronym{CQI}{CQI}{channel quality indicator}
\newacronym{C-RAN}{C-RAN}{Cloud RAN}
\newacronym{CS}{CS}{circuit switched}
\newacronym{CSI}{CSI}{channel state information}
\newacronym{CRE}{CRE}{cell range expansion}
\newacronym{D2D}{D2D}{Device-to-Device}
\newacronym{DFT}{DFT}{discrete Fourier transform}
\newacronym{DL}{DL}{Download}
\newacronym{EFA}{EFA}{Exploratory Factor Analysis}
\newacronym{e2e}{e2e}{End-to-end}
\newacronym{eICIC}{eICIC}{enhanced inter-cell interference cancellation}
\newacronym{eMBB}{eMBB}{enhanced Mobile Broadband}
\newacronym{eNodeB}{eNodeB}{evolved Node-B}
\newacronym{EPC}{EPC}{Evolved Packet Core}
\newacronym{EPS}{EPS}{Evolved Packet System}
\newacronym{E-UTRAN}{E-UTRAN}{Evolved Universal Terrestrial Radio Access Network}
\newacronym{ETSI}{ETSI}{European Telecommunications Standards Institute}
\newacronym{FDMA}{FDMA}{frequency division multiple access}
\newacronym{FFT}{FFT}{fast Fourier transform}
\newacronym{GGSN}{GGSN}{Gateway GPRS Support Node}
\newacronym{GPS}{GPS}{global positioning system}
\newacronym{GRA}{GRA}{Grey relational analysis}
\newacronym{GSM}{GSM}{Global System for Mobile Communications}
\newacronym{GTP}{GTP}{GPRS Tunneling Protocol}
\newacronym{HDFS}{HDFS}{Hadoop Distributed File System}
\newacronym{HetNet}{HetNet}{heterogeneous network}
\newacronym{HiveQL}{HiveQL}{Hive Query language}
\newacronym{HSS}{HSS}{Home Subscriber Station}
\newacronym{HTTP}{HTTP}{Hypertext Transfer Protocol}
\newacronym{ICIC}{ICIC}{inter-cell interference cancellation}
\newacronym{ICN}{ICN}{information-centric network}
\newacronym{IoT}{IoT}{Internet of Things}
\newacronym{IEEE}{IEEE}{Institute of Electrical and Electronics Engineers}
\newacronym{IMEI}{IMEI}{International Mobile Station Equipment Identity}
\newacronym{IMSI}{IMSI}{International Mobile Subscriber Identity}
\newacronym{IMS}{IMS}{IP Multimedia Subsystem}
\newacronym{IMT-A}{IMT-A}{International Mobile Telecommunications - Advanced}
\newacronym{ITU}{ITU}{International Telecommunication Union}
\newacronym{IP}{IP}{Internet Protocol}
\newacronym{JSON}{JSON}{JavaScript Object Notation}
\newacronym{LAC}{LAC}{location area code}
\newacronym{LTE}{LTE}{Long Term Evolution}
\newacronym{LTE-A}{LTE-A}{Long Term Evolution Advanced}
\newacronym{MADM}{MADM}{Multiple Attribute Decision Making}
\newacronym{MEW}{MEW}{multiplicative exponent weighting}
\newacronym{MEC}{MEC}{Multi-acccess Edge Computing}
\newacronym{MIMO}{MIMO}{multiple-input multiple-output}
\newacronym{ML}{ML}{Machine Learning}
\newacronym{MME}{MME}{Mobility Management Entity}
\newacronym{MMF}{MMF}{Max-Min Fair}
\newacronym{MMSE}{MMSE}{minimum mean square error}
\newacronym{MANO}{MANO}{Management and Orchestration}
\newacronym{MNO}{MNO}{Mobile Network Operator}
\newacronym{MVNO}{MVNO}{Mobile Virtual Network Operator}
\newacronym{MSISDN}{MSISDN}{Mobile Station International Subscriber Directory Number}
\newacronym{MTC}{MTC}{Machine Type Communications}
\newacronym{mMTC}{mMTC}{Massive Machine Type Communications}
\newacronym{MT}{MT}{Maximum Throughput}
\newacronym{NB}{NB}{North Bound}
\newacronym{NFV}{NFV}{Network Function Virtualization}
\newacronym{NoSQL}{NoSQL}{Not Only SQL}
\newacronym{NS}{NS}{Network Slice}
\newacronym{OAM}{OAM}{Operation, Administration and Management}
\newacronym{OFDM}{OFDM}{orthogonal frequency division multiplexing}
\newacronym{OFDMA}{OFDMA}{orthogonal frequency division multiple access}
\newacronym{OS}{OS}{operating system}
\newacronym{OSS}{OSS}{Operations Support System}
\newacronym{OTT}{OTT}{over-the-top}
\newacronym{PCRF}{PCRF}{Policy and Charging Rules Function}
\newacronym{PDN}{PDN}{packet data network}
\newacronym{PF}{PF}{Proportional Fair}
\newacronym{PGW}{P-GW}{Packet Data Gateway}
\newacronym{PHY}{PHY}{physical layer}
\newacronym{PPP}{PPP}{{P}oisson point process}
\newacronym{QoE}{QoE}{quality-of-experience}
\newacronym{QoS}{QoS}{quality-of-service}
\newacronym{PDF}{PDF}{probability density function}
\newacronym{PS}{PS}{packet switched}
\newacronym{RAN}{RAN}{Radio Access Network}
\newacronym{RB}{RB}{resource block}
\newacronym{RE}{RE}{range extension}
\newacronym{RF}{RF}{radio frequency}
\newacronym{RG}{RG}{rate guarantee}
\newacronym{RR}{RR}{Round Robin}
\newacronym{RRH}{RRH}{remote radio head}
\newacronym{RRM}{RRM}{radio resource management}
\newacronym{RSSI}{RSSI}{Received Signal Strength Indicator}
\newacronym{RSRP}{RSRP}{reference signal received power}
\newacronym{SAC}{SAC}{service area code}
\newacronym{SAW}{SAW}{simple additive weighting}
\newacronym{SC-FDMA}{SC-FDMA}{single carrier frequency division multiple access}
\newacronym{SCN}{SCN}{small cell network}
\newacronym{SDN}{SDN}{Software Defined Networking}
\newacronym{SDS}{SDS}{Software Defined Storage}
\newacronym{SDO}{SDO}{Standards Developing Organisation}
\newacronym{SGSN}{SGCN}{Serving GPRS Support Node}
\newacronym{SGW}{S-GW}{Serving Gateway}
\newacronym{SHARING}{SHARING}{Self-organized Heterogeneous Advanced RadIo Networks Generation}
\newacronym{SLA}{SLA}{Service Level Agreement}
\newacronym{SINR}{SINR}{signal-to-interference-plus-noise ratio}
\newacronym{SISO}{SISO}{single-input single-output}
\newacronym{SSID}{SSID}{Service Set Identification}
\newacronym{ST}{ST}{Standart Multi-User TOPSIS}
\newacronym{STBCs}{STBCs}{space-time block codes}
\newacronym{SVD}{SVD}{singular value decomposition}
\newacronym{TDMA}{TDMA}{time division multiple access}
\newacronym{TEID}{TEID}{tunnel endpoint identifier}
\newacronym{TOPSIS}{TOPSIS}{Total Order Preference By Similarity to the Ideal Solution}
\newacronym{TTI}{TTI}{transmission time interval}
\newacronym{UE}{UE}{user equipment}
\newacronym{UHD}{UHD}{ultra-high definition}
\newacronym{UL}{UL}{Upload}
\newacronym{UMTS}{UMTS}{Universal Mobile Telecommunications Service} 
\newacronym{URLLC}{URLLC}{Ultra-low reliability and low latency} 
\newacronym{VIM}{VIM}{Virtual Infrastructure Manager}
\newacronym{VNF}{VNF}{Virtual Network Function}
\newacronym{VoIP}{VoIP}{voice over IP}
\newacronym{VoLTE}{VoLTE}{Voice over LTE}
\newacronym{VR}{VR}{Virtual Reality}
\newacronym{W-CDMA}{W-CDMA}{Wideband Code Division Multiple Access}
\newacronym{WiFi}{WiFi}{Wireless Fidelity}
\newacronym{Wi-Fi}{Wi-Fi}{Wireless Fidelity}
\newacronym{WiMAX}{WiMAX}{Worldwide Interoperability for Microwave Access}
\newacronym{WLAN}{WLAN}{Wireless Local Area Network}
\newacronym{WMC}{WMC}{weighted Markov chain}
\newacronym{ZF}{ZF}{zero-forcing}
\newacronym{HO}{HO}{handover}
\newacronym{CapEx}{CapEx}{capital expenditure}
\newacronym{OpEx}{OpEx}{operating expenditure}
\newacronym{BTS}{BTS}{Base Transceiver Station}
\newacronym{PSC}{PSC}{Primary Scrambling Code}
\newacronym{PCI}{PCI}{Physical Cell Identity}
\newacronym{RSI}{RSI}{RACH Root Sequence Index}
\newacronym{RACH}{RACH}{random access channel}
\newacronym{HS-DSCH}{HS-DSCH}{High Speed Downlink Shared Channel}
\newacronym{PDCP}{PDCP}{Packet Data Convergence Control}
\newacronym{NAS}{NAS}{Non Access Stratum}
\newacronym{KPI}{KPI}{key performance indicator}
\newacronym{IT}{IT}{information technology}

\begin{document}

\title{An Experimental Study of Factor Analysis  over Cellular Network Data }


\author{Feyzullah Kalyoncu, Engin Zeydan, Ahmet~Yildirim and  Ibrahim Onuralp Yigit\\
T\"{u}rk Telekom Labs, Istanbul, Turkey 34889\\
E-mail: \{feyzullah.kalyoncu, engin.zeydan, ibrahimonuralp.yigit\}@turktelekom.com.tr, ahmet.yildirim3@boun.edu.tr\\}

\maketitle

\begin{abstract}
Mobile Network Operators (MNOs) are evolving towards becoming data-driven, while delivering capacity to collect and analyze data. This can help in enhancing user experiences while empowering the operation workforce and building new business models. Mobile traffic demands of users can give insights to MNOs to plan, decide and act depending on network conditions. In this paper, we investigate the behaviour of Istanbul residents using the cellular network traffic activity over spatial and temporal dimensions via exploratory factor analysis (EFA) using a major MNO's cellular network traffic data in Turkey. Our results reveal various time and spatial patterns for Istanbul residents such as morning and evening commuting factors, business and residential factors as well as nightlife and weekend afternoon factors as the most prominent cultural behaviour. The analysis results also demonstrate interesting findings such as tunnels and transportation paths selected by Istanbul residents may differ during morning rush work hour compared to evening rush after-work hour.


\end{abstract}


\IEEEpeerreviewmaketitle

\section{Introduction}

In the era of digital age, data can inform and empower transformations of all industry such as telecommunications, health, automotive and factories of future.  Using the data as a commodity, organization institutions can detect important phenomena as early as possible, can forecast the outcomes that are yet to come (predictive) or can perform optimization for better modeling and outcomes (prescribe).  Some of the enablers for this kind of process enhancement and gaining  enterprise level intelligence from rich dataset are via machine learning and statistical methodologies~\cite{naboulsi2016large}.  The cellular data collected from various locations at different time provides rich and important context data  to \glspl{MNO} for better decision making and  operational process enhancements. For example, the collected data can provide insights into different mobility patterns for groups of subscribers over the observed duration periods after data analysis. This can enhance the capabilities of \glspl{MNO} for providing personalized services for different customer segments based on their personalized experiences. The mobile data interaction can reflect the personal behaviour, preferences and objectives of users. For this reason, \glspl{MNO} are working to improve operational efficiency, obtain predictive analytics and extract insights by analyzing massive dataset available in their premises. 








For controlling and optimization of network operations, historical and contextual data analysis for modeling the traffic at cell level such as~\cite{paul2011understanding,lu2013approaching},\cite{song2010limits,narmanliogluprediction} and profiling the user behaviour at different time scales such as~\cite{furno2017joint,naboulsi2014classifying} based on mobile traffic demand exist in the literature. An analysis that relies on \ac{EFA} using with real-world mobile traffic dataset for Milan and Paris cities are performed in~\cite{furno2017joint}. The results reveal different network activity profiles in those two major European cities. In this paper, we further extend the \ac{EFA} presented for Milan and Paris cities and study network activity profiles of Istanbul using real-world mobile traffic dataset of a major \ac{MNO} in Turkey. Compared to previous works in mobile data analysis, our results  reveal various mobile usage demands of Istanbul residents and also shed some light on  cultural behaviour. 

\section{System Model and Architecture}

\begin{figure}[htp!]
\centering
\includegraphics[width=1.0\linewidth]{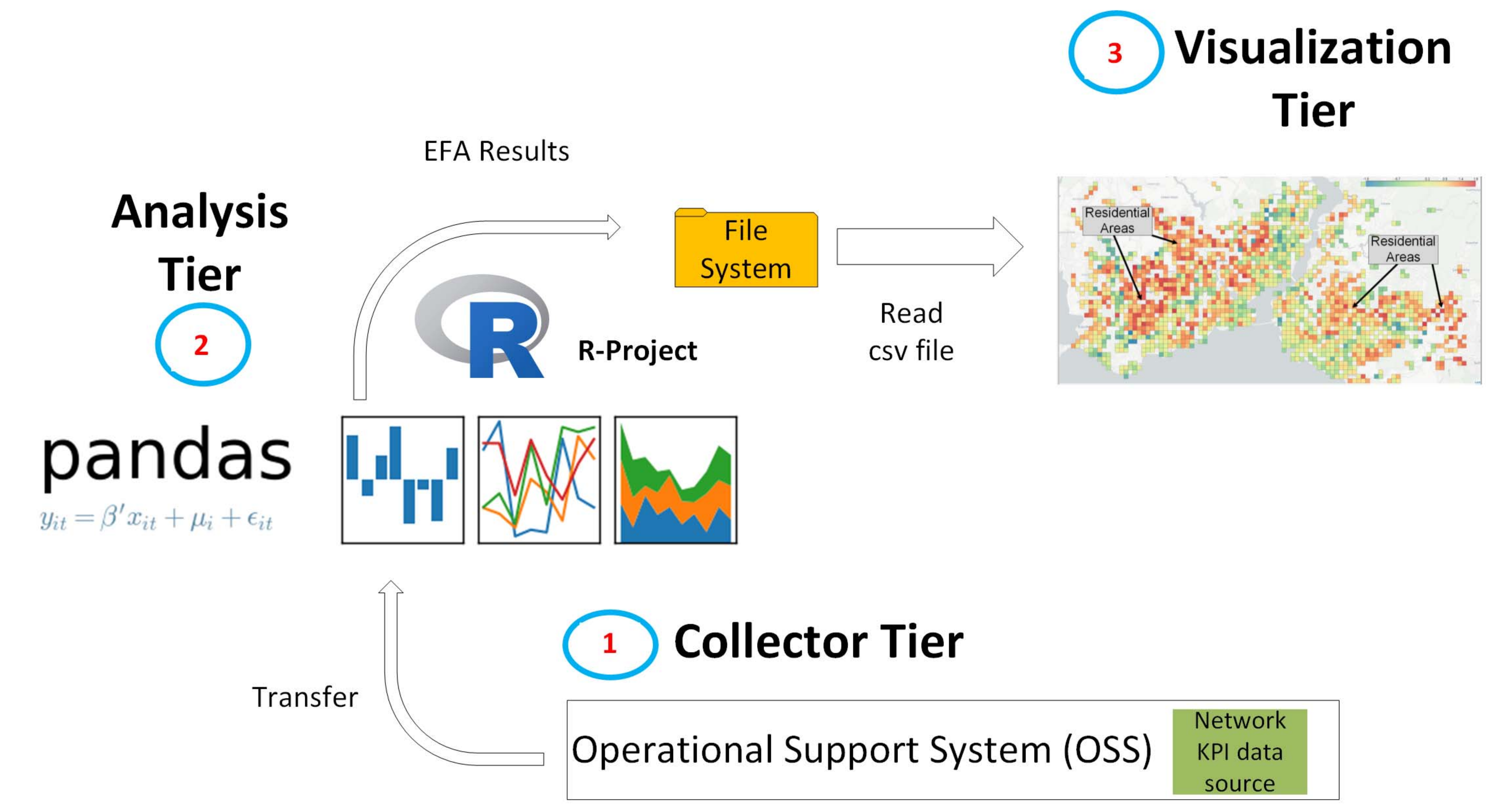}
\caption{System Architecture for EFA and data visualization}
\label{fig:SIGMONA_arch}
\end{figure}

The proposed platform processes and analyzes network \ac{KPI} data that is collected a priori from \ac{OSS} of the \ac{MNO} in Turkey. Fig.~\ref{fig:SIGMONA_arch} demonstrates the general architecture of our utilized solution.  The proposed architecture is composed of three main modules:  \textbf{(a) Data Collector Tier},  \textbf{(b) Analysis Tier},  \textbf{(e) Visualization Tier}.  The solution is based on utilization of the network \ac{KPI} data  together with open-source data analytics software and platforms.   \textit{Pandas} and \textit{R} packages are used as data analysis tools utilizing data-centric packages.  For map visualization, Folium visualization tool~\cite{Folium} and inside the Folium, Leaflet javascript~\cite{leaflet} library are utilized. 

\begin{figure}[htp!]
\centering
\includegraphics[width=0.8\linewidth]{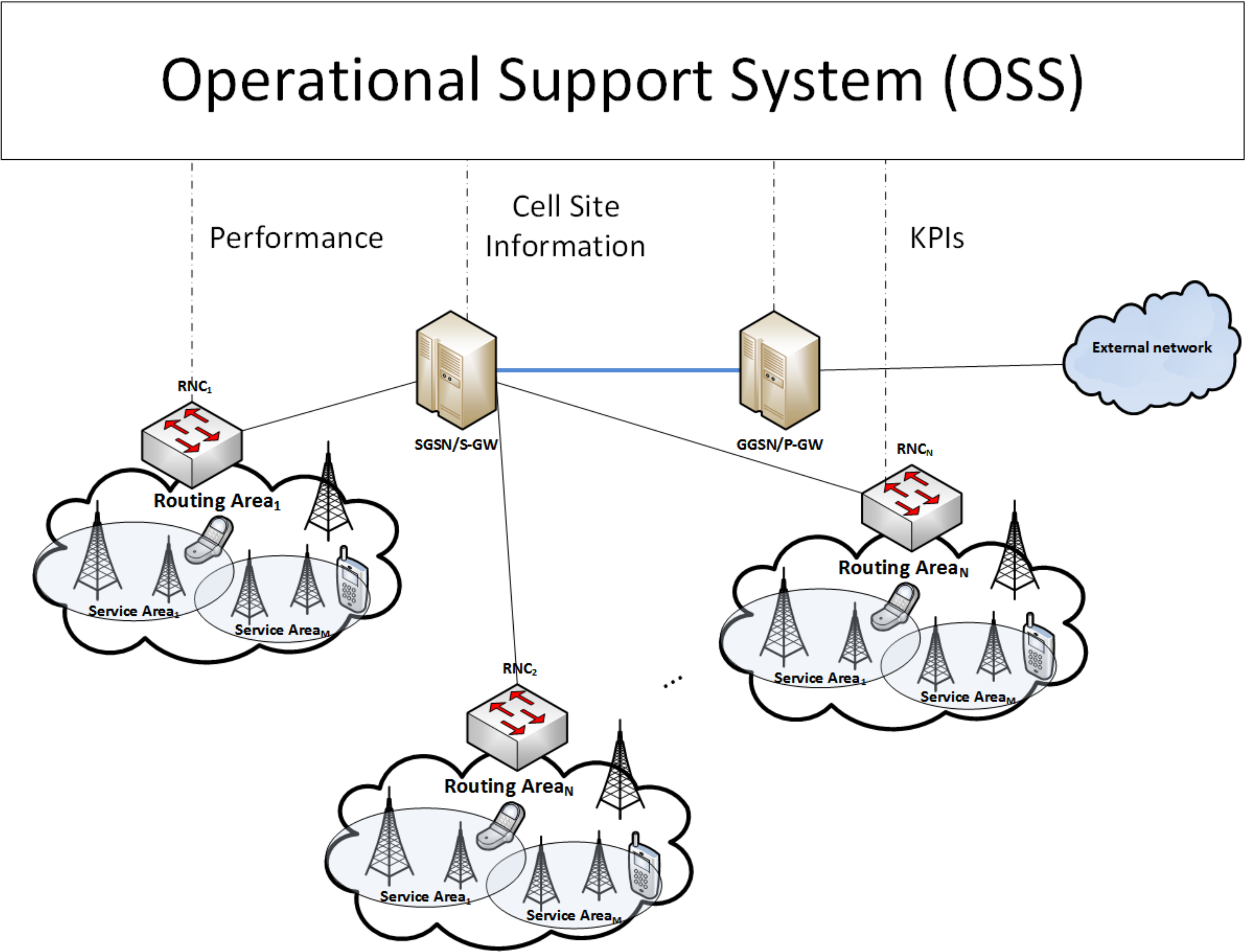}
\caption{An illustrative architecture for cellular data collection in \ac{OSS}.}
\label{fig:test_database}
\end{figure}

In \textbf{Data Collector Tier} marked as step-(1) in Fig.~\ref{fig:SIGMONA_arch}, the data is collected from \ac{OSS} similar to Fig.~\ref{fig:test_database} and is transferred into Pandas and R data analytics tools marked as step-(2).  In Pandas, first the data in csv format is grouped by Site-ID and the latitude, longitude of the sites are appended into the existing csv data format. In step-(2), all the \ac{PS} traffic corresponding to each Cell-ID for 4G  are analyzed.  An example  of sum 4G traffic values for Besiktas and Umraniye District of Istanbul is given in Table~\ref{cell_site_evaluation}. Finally, a matrix with each Cell-ID as row and hourly median \ac{PS} traffic values over a week in 4G as column matrix is constructed and fed into \ac{EFA} for determining different factors from the cellular data traffic. After \ac{EFA}, different factors and corresponding \ac{BS} scores for each factor are obtained. Later, the anaysis results are visualized in \textbf{Visualization Tier},  marked as step-(3) in Fig.~\ref{fig:SIGMONA_arch}, where we utilize Folium \& Leaflet maps in order to visualize the \ac{EFA} scores of each cell sites for each obtained factor. 


\textbf{Model:} Let \textbf{X} be a $N \times 1$ vector of observed variables, i.e. phenomena of interest such as \ac{DL}, \ac{UL} traffic or number of users on a given \ac{BS} in this paper. The fundamental equation for factor analysis is defined as~\cite{furno2017joint}

\begin{equation}
    \textbf{X} = \textbf{A} \textbf{F} + \textbf{U} 
    \label{weighted_combination}
\end{equation}

\noindent where \textbf{A} represents $N \times K$ matrix of common factor pattern coefficients that describe importance of each factor to every variable, \textbf{F} represents $K \times 1$ vector of unknown normalized common factors, i.e., a small number (K$<<$N) of complex relationships among variables and \textbf{U} represents $N \times 1$ vector of unknown unique factors that are specific to a single variable. Hence, \eqref{weighted_combination} is a weighted combinations of the common factors in \textbf{F} and the unique factors in \textbf{U}. Together with \ac{EFA} solution, by analyzing variable observations  from a set of samples, \ac{EFA} can identify common/unique factors, and numerical relationships that describe how much each common factor explains each variable~\cite{mulaik2009foundations}.

In our analysis, in order to determine the number of factors to extract,  we have utilized \textit{parallel analysis} where the largest eigenvalues of the data correlation matrix is selected. For factor rotation, we have selected \emph{promax} for maximizing the high loadings using R project~\cite{R_project}.

\begin{table}[h]
\centering
\small
\caption{An example  of sum 4G traffic values for Besiktas and Umraniye District of Istanbul.}
\label{cell_site_evaluation}
\begin{tabular}{|l|l|l|l|l|}
\hline
\textbf{District} & \makecell{\textbf{\ac{DL} Traffic} \\ \textbf{(GB)}} & \makecell{\textbf{\ac{UL} Traffic} \\ \textbf{(GB)}} & \makecell{\textbf{No. of Users} \\ \textbf{(avg.)}} \\ \hline 
Besiktas  & 4,239,906.386 & 488,613.8446 & 67,460.33 \\ \hline
Umraniye  & 9,460,358.224 & 870,681.5689 & 142,626.2021 \\ \hline
\end{tabular}
\end{table}

\begin{table}[h]
\centering
\small
\caption{Statistics of Analyzed Cellular Data  in Istanbul.}
\label{TestStatistics}
\begin{tabular}{|l|l|}
\hline
\textbf{\# of rows}                       &  6,264,286 \\ \hline 
\textbf{\# of districts}                        &   40 \\ \hline
\textbf{PS \ac{DL} traffic (average per day, Gb)}     &     18,318.87544 \\ \hline
\textbf{PS \ac{UL} traffic (average per day, Gb)}                  &    1,793.451409   \\ \hline
\textbf{Obser. Duration}                  &   1 month   \\ \hline
\textbf{Average \# of active users}                  &    284.5   \\ \hline
N, K                  &    7 $\times$ 24, 10-13   \\ \hline
\end{tabular}
\end{table}

\begin{table*}[ht]
\centering
\small
\caption{Factors Analysis Descriptions in Istanbul}
\label{Land_Use}
\begin{tabular}{|c|l|l|}
\hline
\textbf{Factor} & \multicolumn{1}{c|}{\textbf{Labeled Areas}} & \multicolumn{1}{c|}{\textbf{Description}} \\ \hline
\textbf{DL 1}      & \makecell{Residential Areas}                              &      \makecell{Large  population residential areas  in western part of \\ European side and eastern part of Anatolian side}                                 \\ \hline
\textbf{No. of Users 1}      & \makecell{Office, Campus and \\ Industrial Areas}                             & \makecell{University campuses and business zones in Maslak, industrial \\  areas around Basaksehir and commercial zones around Besiktas, Fatih  \\  in European side and Uskudar and Kadikoy (commercial regions) and  \\  Dudullu (industrial regions) in Anatolian side   }                                     \\ \hline
\textbf{No. of Users 2}      & \makecell{Malls, Touristic Areas \\  and Leisure Activity}       & \makecell{Historical Regions (Sultanahmet, Kapalicarsi, Galata tower), \\ Shopping centers in Maslak and Bakirkoy, \\ Leisure time activity on coastal side and Boshoporus view sites}                                      \\ \hline
\textbf{DL 2}      & \makecell{Morning Commuting}                                      & \makecell{Commuting traffic from Anatolian side into European side over bridges \\ Metrobus line in European side and Avrasya tunnel entrance and exit points}                                      \\ \hline
\textbf{DL 3}      & \makecell{Evening Commuting}                                      & \makecell{Metrobus express bus line in European side and \\ Highway hub in Anatolian side  }                                       \\ \hline
\textbf{DL 4}      & \makecell{Farmer's market, Major  \\ Bus Terminal and Airport, \\ nightlife area}                                         & \makecell{Wholesale food market  and \\ Esenler Bus Major Terminal in Bayrampasa  \\ Ataturk airport and nightlife (Taksim, Besiktas area)}                                      \\ \hline
\end{tabular}
\end{table*}


\section{Network factoring based on traffic usage}
\label{performance}

Using the spatio-temporal characteristics of cellular data, we investigate the performance of utilizing \ac{EFA} over different considered variables, i.e. DL, UL and number of users at each \glspl{BS}. Our focus is on summarized data on a period of one-week and 24 hours that can represent cellular traffic data collected over a one month period. For this reason, similar to~\cite{furno2017joint}, one month data is condensed into one single median week, i.e. the median of all measurements are obtained for each hour and day of the week for each site in Istanbul. 

\textbf{Dataset:} Our dataset consists of cellular network traffic in Istanbul of a major \ac{MNO} in Turkey.  The statistical parameters of the utilized data is given in Table~\ref{TestStatistics}.  In our \textit{Dataset} there exists \ac{PS} traffic (both data and \ac{VoLTE}) from geographically distributed region in the country dated from 29 November 2017 to 26 December 2017. The data fields we have utilized consist of hourly average of \ac{DL} and \ac{UL} traffic data and number of active \glspl{UE} for  \ac{PS} data in 4G network infrastructures as well as date, region, city, district, Site-ID and Cell-ID information of each \ac{BS}. Total number of rows in the data is 6,264,286. Hence, the mobile traffic is large enough to perform statistical significant studies.

\begin{figure*} [ht!]
\begin{subfigure}{.5\textwidth}
  \centering
  \includegraphics[width=\linewidth]{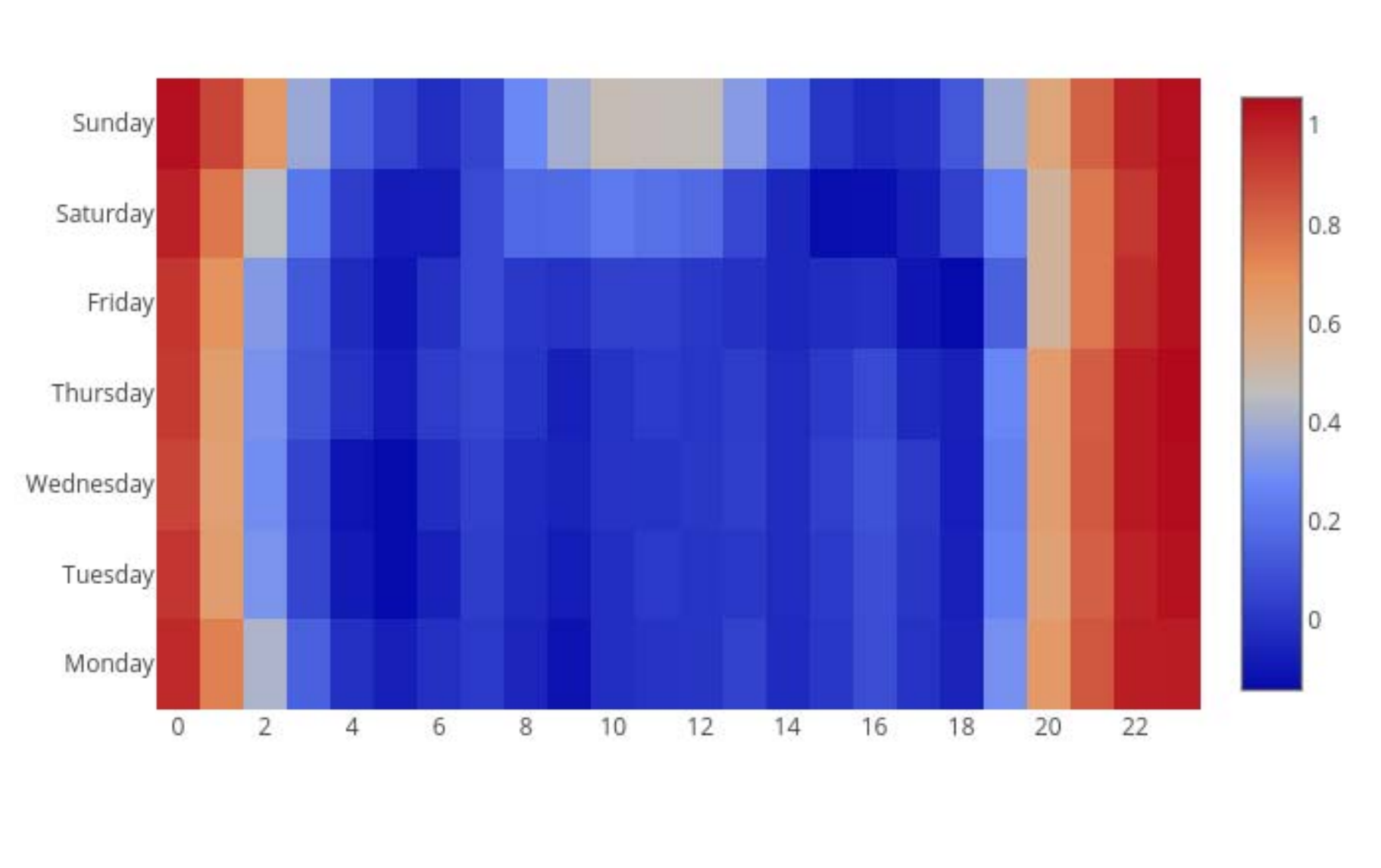}
  \caption{}
  \label{factor_1}
\end{subfigure}%
\begin{subfigure}{.5\textwidth}
  \centering
  \includegraphics[width=\linewidth]{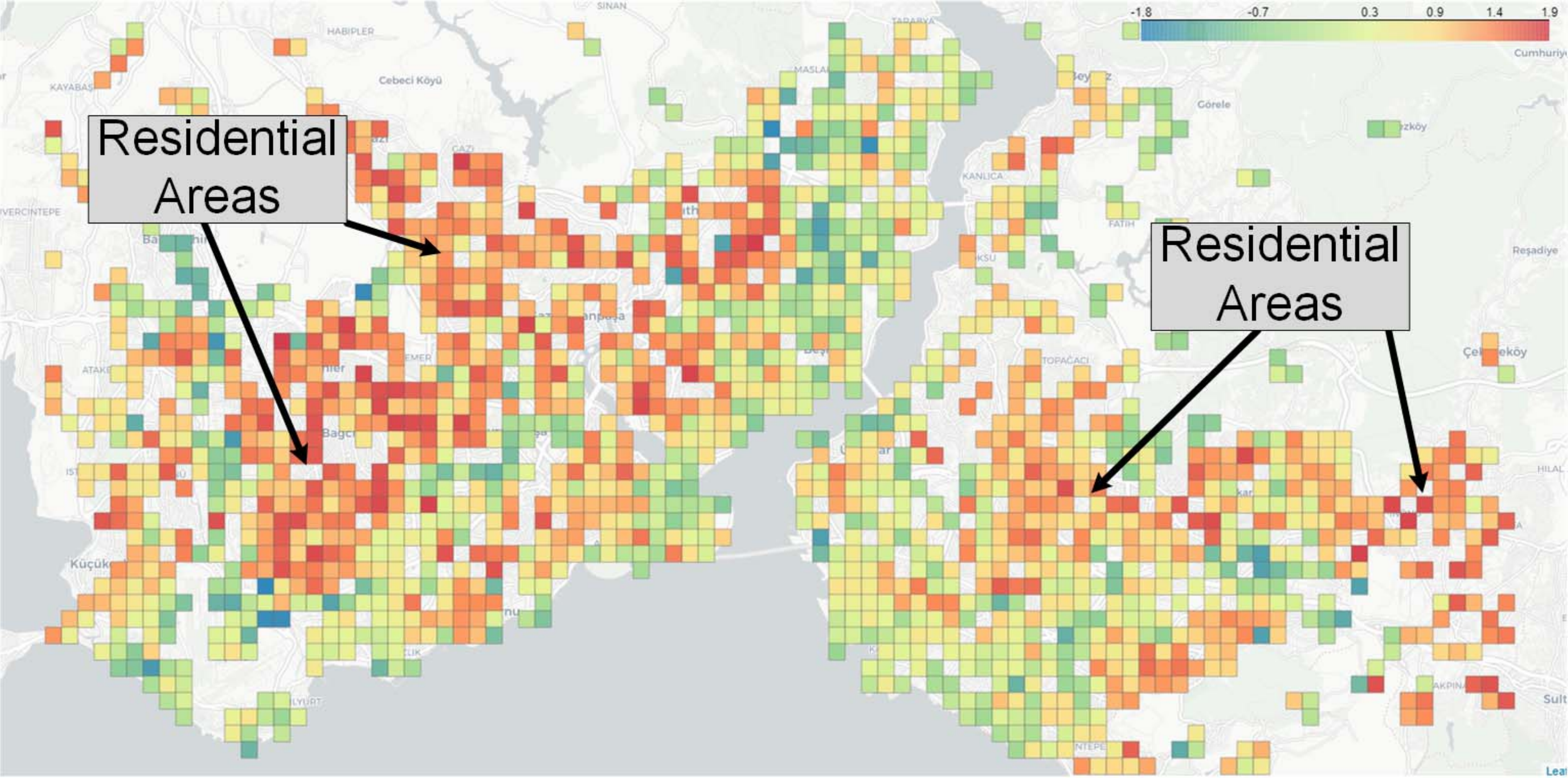}
  \caption{}
  \label{factor_1_map}
\end{subfigure} 
\begin{subfigure}{.5\textwidth}
  \centering
  \includegraphics[width=\linewidth]{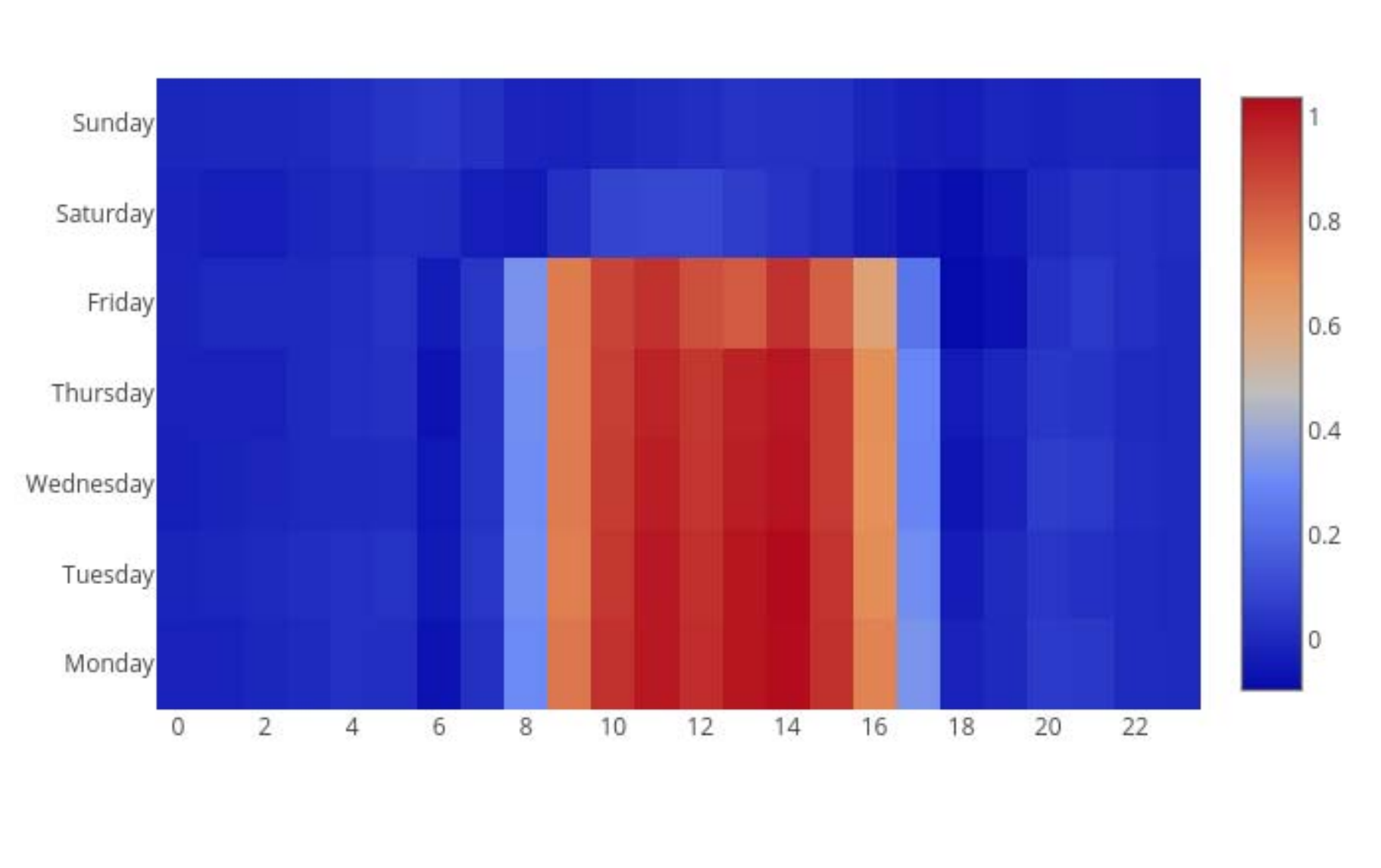}
  \caption{}
  \label{factor_1_user}
\end{subfigure}
\begin{subfigure}{.5\textwidth}
  \centering
  \includegraphics[width=\linewidth]{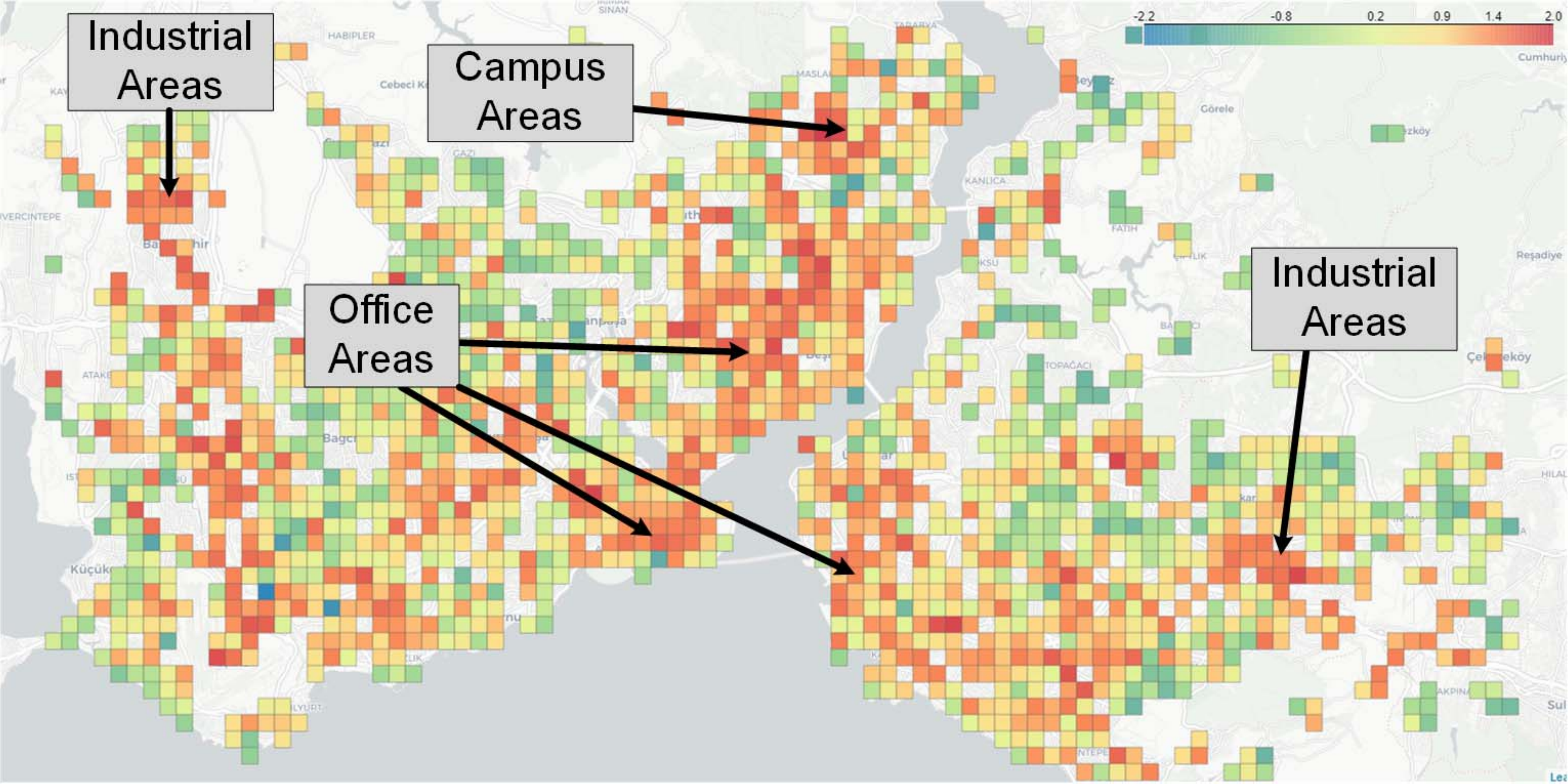}
  \caption{}
  \label{factor_1_user_map}
\end{subfigure}
\begin{subfigure}{.5\textwidth}
  \centering
  \includegraphics[width=\linewidth]{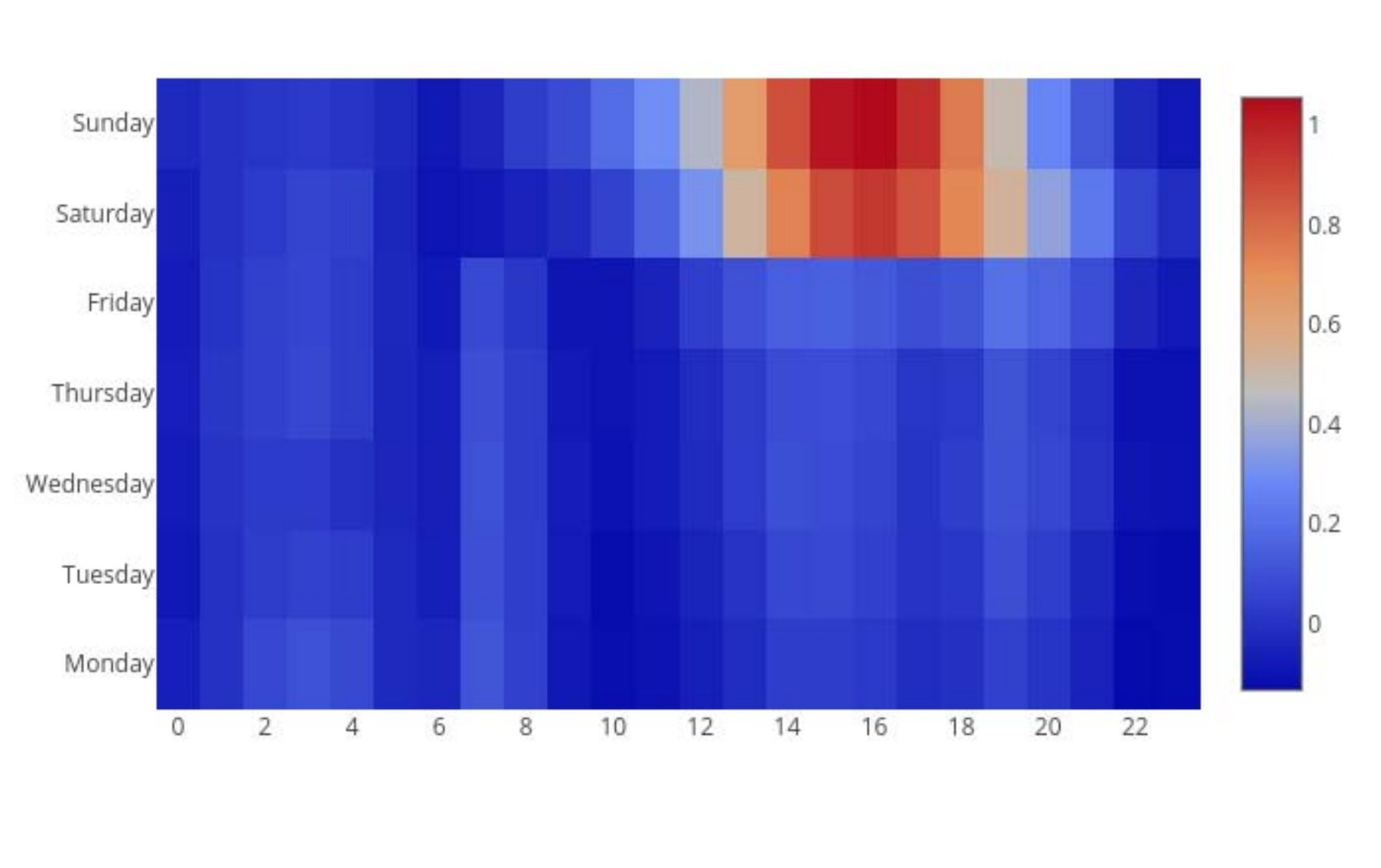}
  \caption{}
  \label{factor_2_user}
\end{subfigure}
\begin{subfigure}{.5\textwidth}
  \centering
  \includegraphics[width=\linewidth]{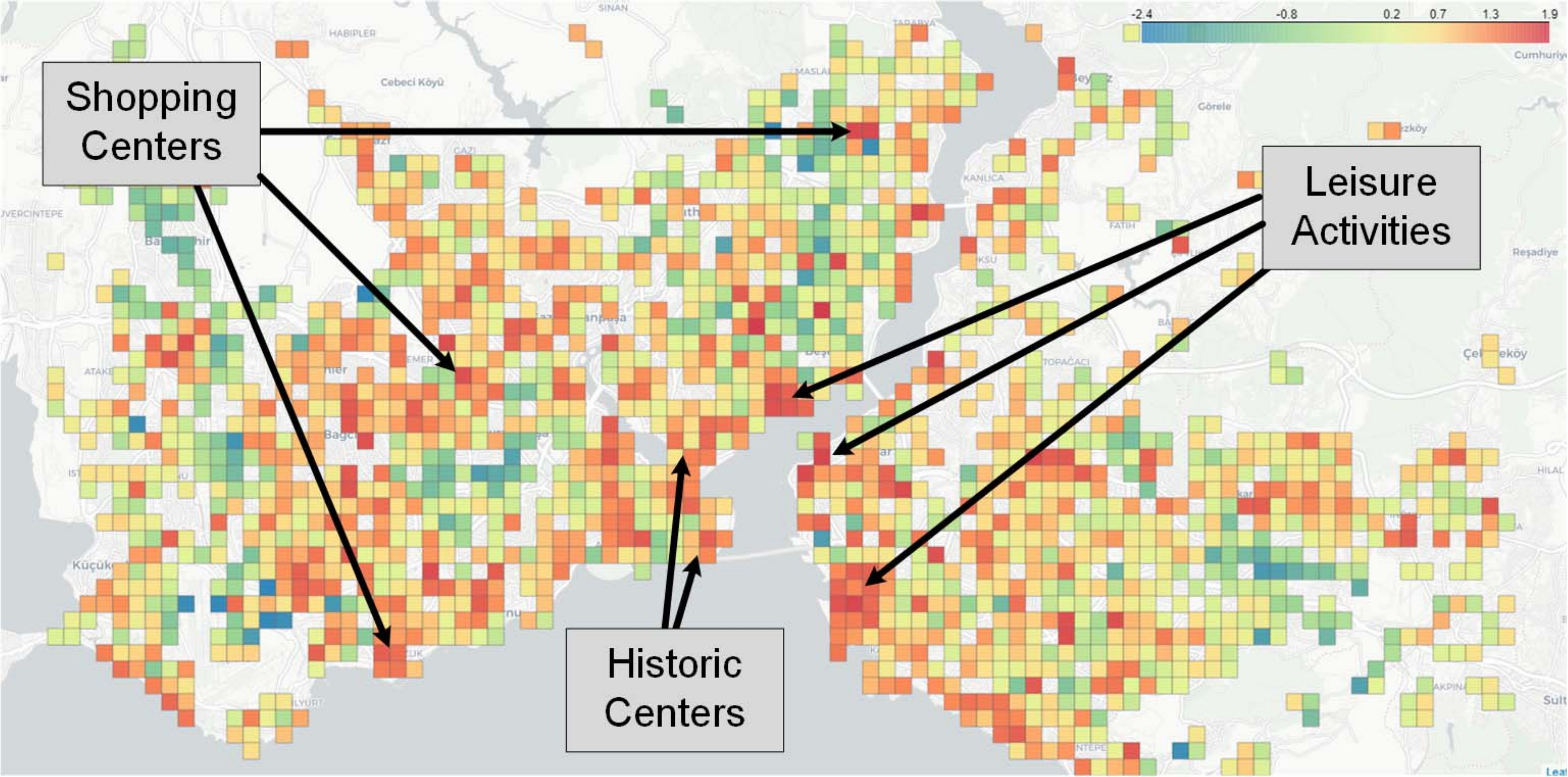}
  \caption{}
  \label{factor_2_user_map}
\end{subfigure}
\caption{ (a) DL Factor 1  (b) DL Factor 1 Map   (c)  Number of Users Factor 1 (d) Number of Users Factor 1 Map (e) Number of Users Factor 2 (f) Number of Users Factor 2 Map}
\label{factor_analysis}
\end{figure*}

\begin{figure*} [ht!]
\begin{subfigure}{.5\textwidth}
  \centering
  \includegraphics[width=\linewidth]{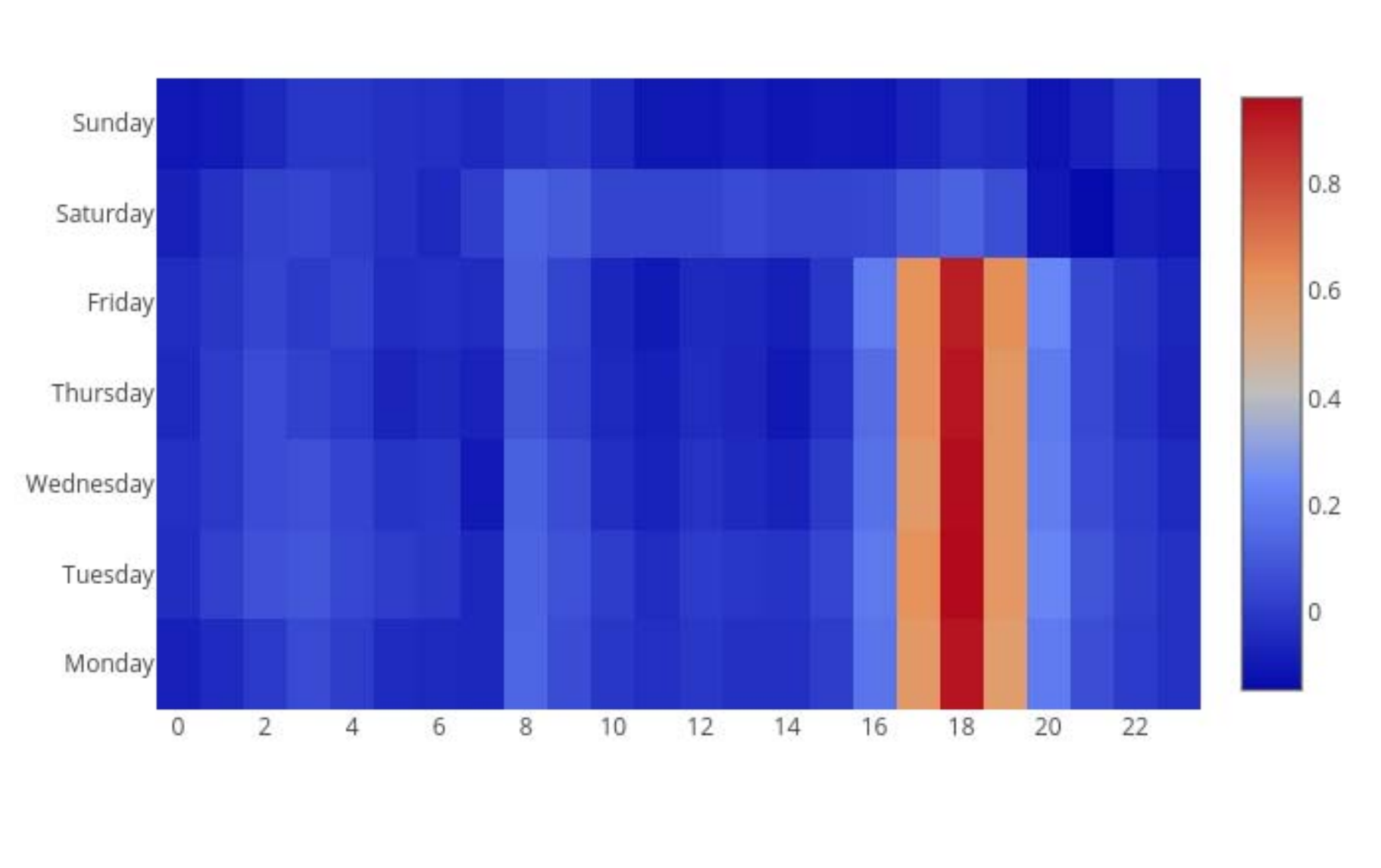}
  \caption{}
  \label{factor_2}
\end{subfigure}%
\begin{subfigure}{.5\textwidth}
  \centering
  \includegraphics[width=\linewidth]{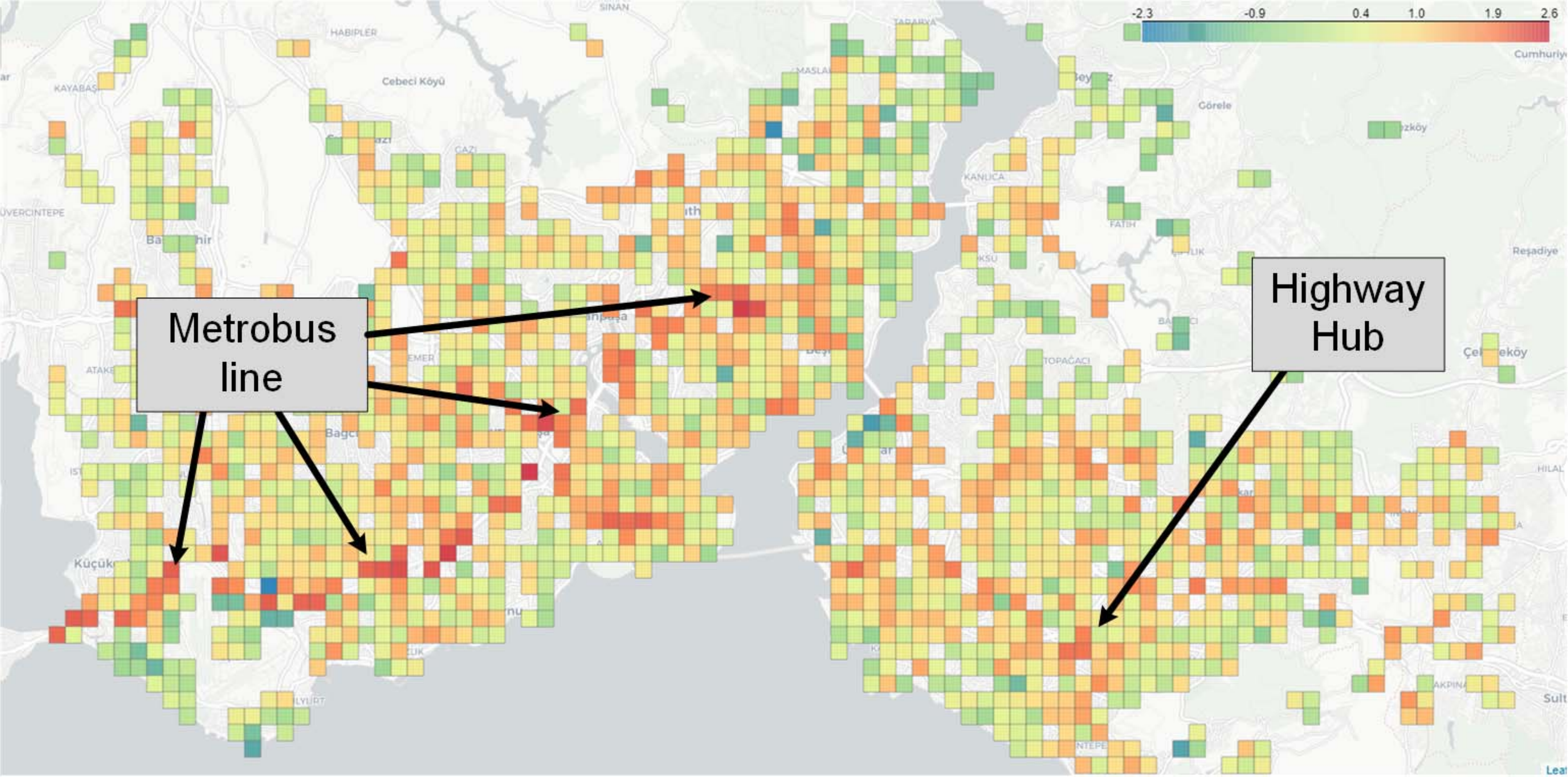}
  \caption{}
  \label{factor_2_map}
\end{subfigure} 
\begin{subfigure}{.5\textwidth}
  \centering
  \includegraphics[width=\linewidth]{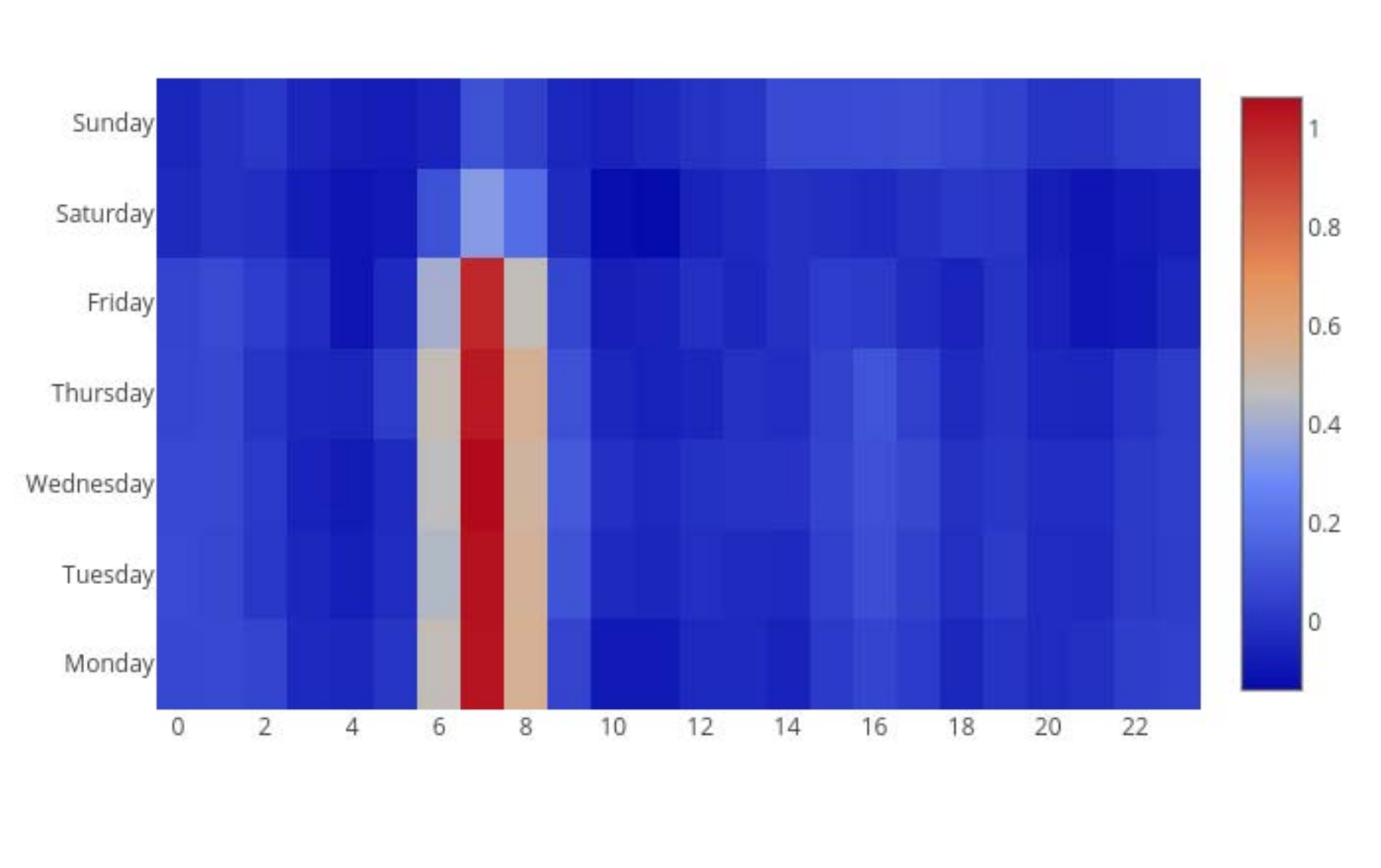}
  \caption{}
  \label{factor_3}
\end{subfigure}
\begin{subfigure}{.5\textwidth}
  \centering
  \includegraphics[width=\linewidth]{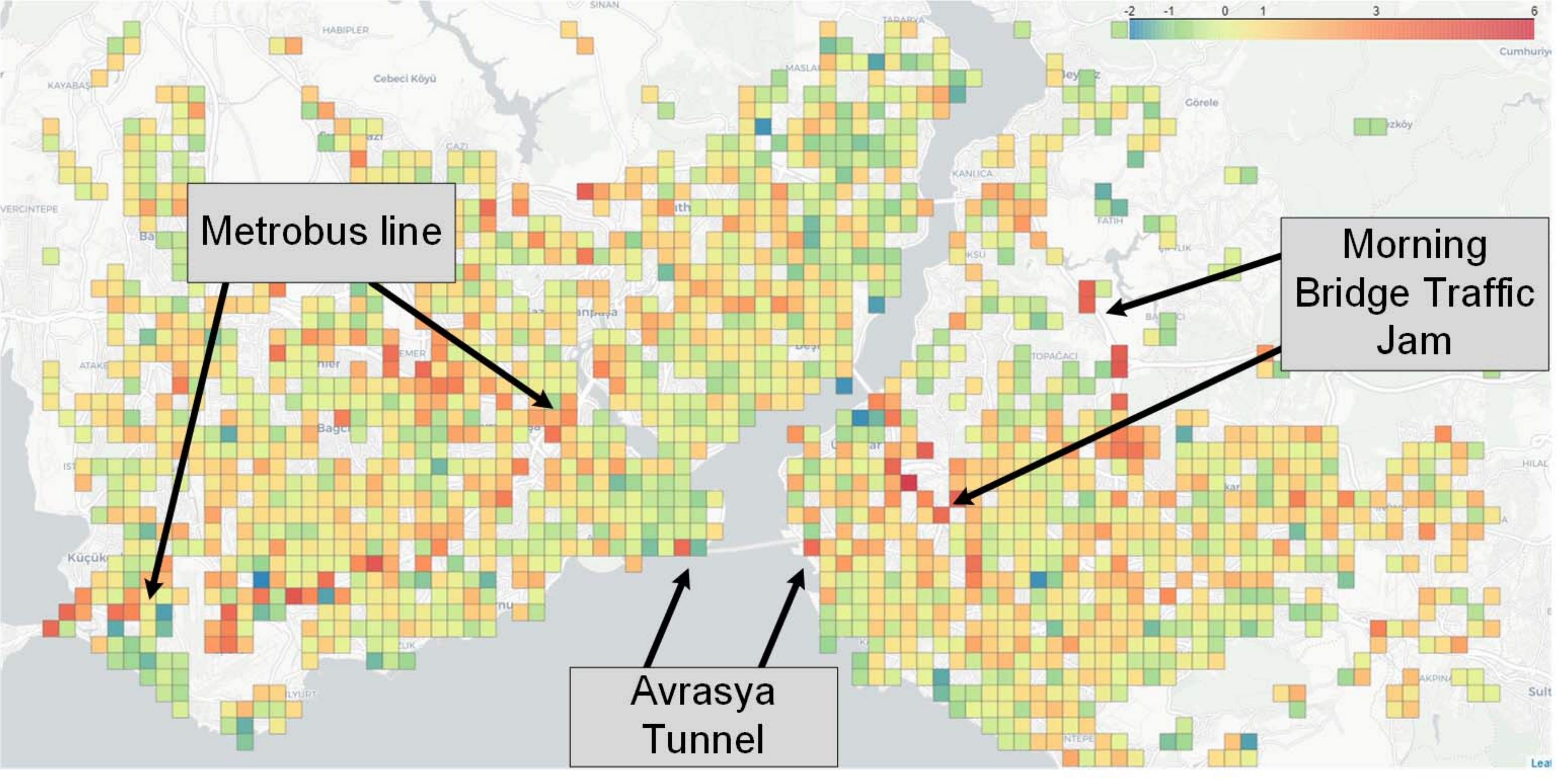}
  \caption{}
  \label{factor_3_map}
\end{subfigure}
\begin{subfigure}{.5\textwidth}
  \centering
  \includegraphics[width=\linewidth]{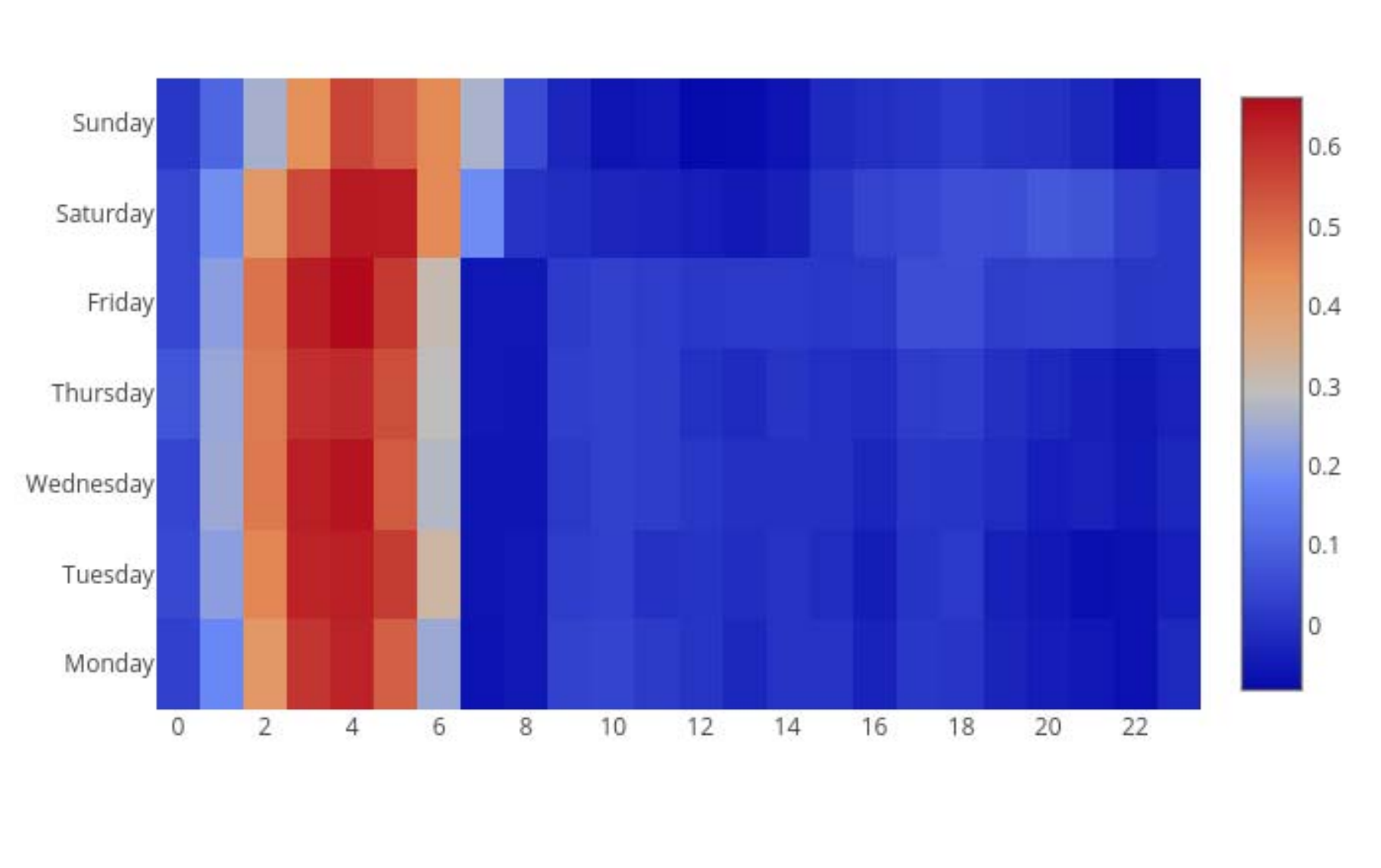}
  \caption{}
  \label{factor_4}
\end{subfigure}
\begin{subfigure}{.5\textwidth}
  \centering
  \includegraphics[width=\linewidth]{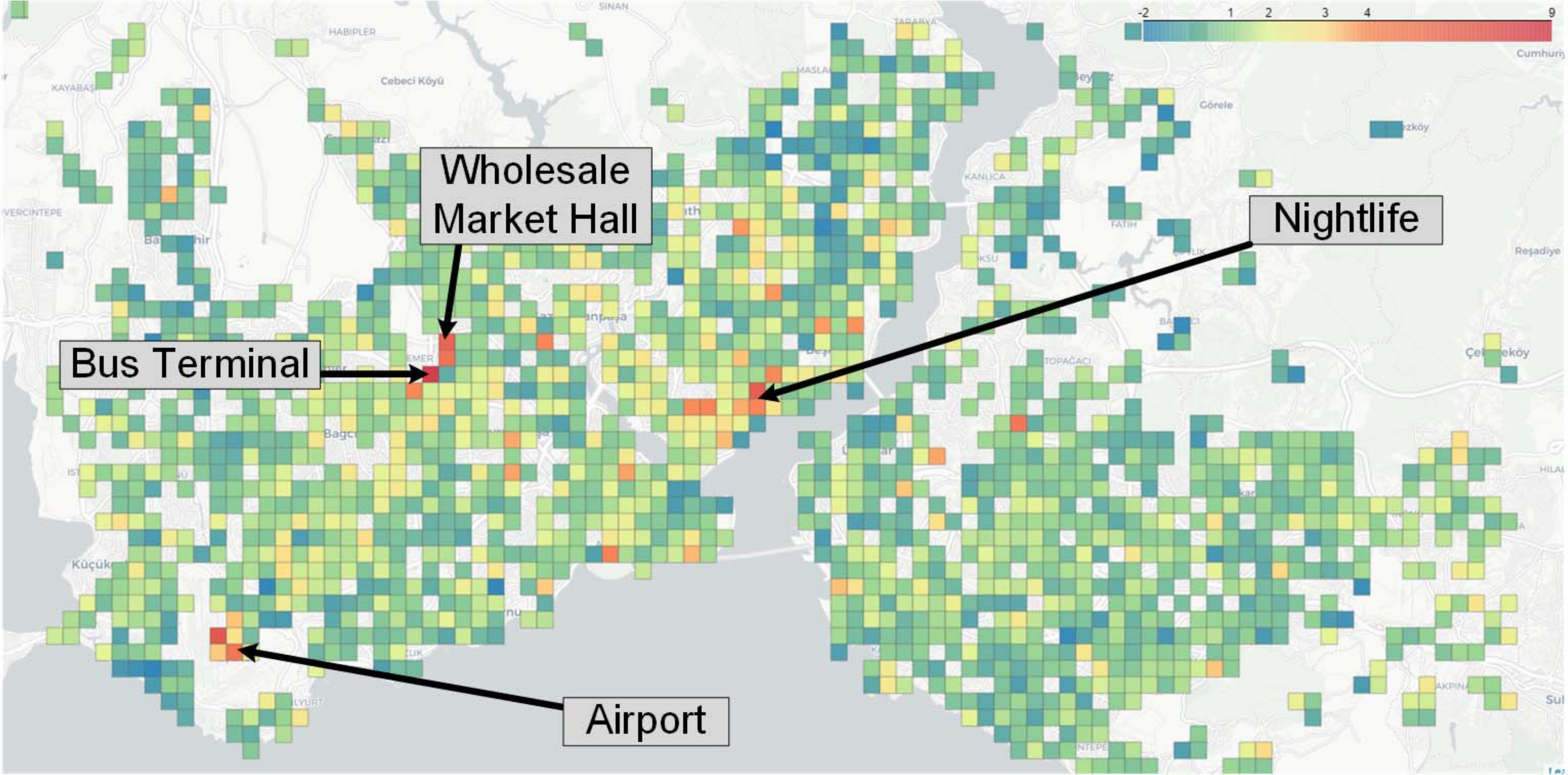}
  \caption{}
  \label{factor_4_map}
\end{subfigure}
\caption{ (a) DL Factor 2  (b) DL Factor 2 Map  (c) DL Factor 3  (d) DL Factor 3 Map  (f) DL Factor 4  (g) DL Factor 4 Map}
\label{factor_analysis_map}
\end{figure*}

\textbf{Interpretation of EFA:}  Using \ac{EFA}, we can obtain different factors which show both temporal and spatial data traffic demands  over various day time and  week days. When we observe the \ac{EFA} scores of each \ac{BS}, we can induce which specific cells belong to each \ac{EFA} factor. After performing \ac{EFA} over our dataset, \emph{optimal} number of different factors, denoted by K, in our analysis  is obtained to be 13 for \ac{UL}, 10 for \ac{DL} and 12 for number of users. For the selected six factors of Fig.~\ref{factor_analysis} and Fig.~\ref{factor_analysis_map}, we have identified different patterns in various regions and time. Other factors have also emerged which indicates some anomalies or special day events which we have excluded due to space limitations.

Identified \ac{EFA} factors are shown in Fig.~\ref{factor_analysis} and Fig.~\ref{factor_analysis_map}. The locations of each cell-IDs over the selected six \ac{EFA} factors of \ac{DL}, \ac{UL} and number of active users are plotted as heatmap using Folium map visualization format. From these figure, we can observe that e.g. for Number of User Factor 1 of Fig.\ref{factor_1_user}, there exists high traffic between 9 am to 6 pm from Monday to Friday, whereas for DL Factor 3 of Fig.\ref{factor_3}, high network utilization exists between 6 am to 9 am.  Table~\ref{Land_Use} explains the observed factor analysis labels and corresponding location descriptions in Istanbul. 
From Fig.~\ref{factor_analysis}, we can observe how the cells belonging to DL Factor-1 of Fig.~\ref{factor_1} and Fig.~\ref{factor_1_map} are highlighting the areas that are mostly populated with local residents, where most of the residential areas are located.   Most of the network activity is around  residential centers such as Bagcilar, Gungoren, Alibeykoy in European side and Umraniye, Sultanbeyli in Anatolian side. Geographical areas where Number of Users Factor 1 as shown in  Fig.~\ref{factor_1_user} and Fig.~\ref{factor_1_user_map}   are mostly related to university campuses, office and industrial areas of Istanbul. This is in line with the properties of  Number of Users Factor 1 of Fig.~\ref{factor_1_user}  which characterize the  working hour traffic between 8 am to 5 pm belonging to \glspl{BS} whose mobile data traffic activity surges during working hours.  Number of Users Factor 2 of Fig.~\ref{factor_2_user} and Fig.\ref{factor_2_user_map}, on the other hand, characterize the touristic, shopping and leisure activity areas in Istanbul. The historic city center Sultanahmet, Kapalicarsi, Galata Tower that are major touristic attractions for foreigners and local tourists, big shopping malls for local residents such as Istinye Park in Maslak (northern European side ), Cevahir and  Zorlu Center in Levent (middle European side), Mall of Istanbul, Ataturk airport shopping center (AVM),  Marmara Forum, Galeria, Forum Istanbul, Ikea, Atakoy Plus (in southern European side) are highlighted in Number of Users Factor 2 of Fig.~\ref{factor_2_user} and Fig.~\ref{factor_2_user_map}. Another dimension in Number of Users Factor 2  that overlaps with the shopping mall activities is indication of the leisure occupations of Istanbul residents around both Anatolian and European coastal sides as well as Boshoporus view regions.

We can also observe different patterns from the other factor profiles. \ac{DL} Factor 2 and \ac{DL} Factor 3 of Fig.~\ref{factor_analysis_map} show the commuting behaviour of Istanbul residents. In Istanbul, most of the residents live in Anatolian side  and commute to European side of Istanbul for work. Therefore,  after work between 5-8 pm of Fig.~\ref{factor_2} and Fig.~\ref{factor_2_map}, there exists coherent usage of public transport services, e.g. Metrobus express bus line used by local commuters is running in major locations of Istanbul can be clearly observed from Fig.~\ref{factor_2_map}). During morning commuting hours between 7 am to 9 am of Fig.~\ref{factor_3} and Fig.~\ref{factor_3_map}, the scores are somehow distributed around entrance from Anatolia side into European side due to existence of non-uniform starting hours of businesses (ranging from 6 am to 9 am) and Metrobus line is less visible compared to evening commuting. In \ac{DL} Factor 3 of Fig.~\ref{factor_3} and Fig.~\ref{factor_3_map},  other active areas that morning commuters frequently use are entrance and exit points of Avrasya tunnel (which runs under Marmara sea between Anatolian side and European side) where morning commuters enter from Anatolian side and exist into European side. We can also notice the interesting fact that commuter behaviour around usage of Avrasya tunnel is only visible in morning commuter behaviour, not in evening commuter behaviour.

The areas highlighted by  DL Factor 4 of Fig.~\ref{factor_4} and Fig.~\ref{factor_4_map} show the activities between 2-4 am where major bus station, major airport, wholesale market hall as well as nightlife areas in Istanbul are highlighted around those times.  The geographical cells that have high scores belonging to DL Factor 4 demonstrate wholesale market hall where all the goods (fruits and vegetables) are stocked for servicing the next day.  DL Factor 4  is also indicating nightlife occupations of Istanbul residents that overlap with terminal (bus and airport)  activities.

\section{Conclusions}

In this paper, we have investigated hourly cellular traffic data and performed factor analysis over the collected dataset of one month of a major cellular mobile network operator in Turkey. The results reveal that there exists different patterns of traffic over different factors depending on the day of the week as well as time of the day. Our results have revealed major residential, business, touristic and market areas as well as morning and evening commuter paths of residents in Istanbul depending on the behaviour of cellular network usage over different time zones. As a future work, we are working on extending the analysis over one year period that can also cover the special events and holidays. 


\bibliographystyle{ieeetr}
\bibliography{References}

\balance

\end{document}